\documentclass[conference]{IEEEtran}
\IEEEoverridecommandlockouts
\usepackage{cite}
\usepackage{amsmath,amssymb,amsfonts}
\usepackage{algorithmic}
\usepackage{graphicx}
\usepackage{textcomp}
\usepackage{xcolor}
\usepackage{eurosym}

\def\BibTeX{{\rm B\kern-.05em{\sc i\kern-.025em b}\kern-.08em
    T\kern-.1667em\lower.7ex\hbox{E}\kern-.125emX}}

\begin{document}


\title{
Modern Techniques for Ancient Games
\thanks{Funded by a \euro2m European Research Council (ERC) Consolidator Grant.}
}

\author{
\IEEEauthorblockN{Cameron Browne}
\IEEEauthorblockA{{\it Games and AI Group} \\ {\it Department of Data Science and Knowledge Engineering (DKE)} \\
{\it Maastricht University} \\
Maastricht, The Netherlands \\
cameron.browne@maastrichtuniversity.nl}
}

\maketitle


\begin{abstract}
Games potentially provide a wealth of knowledge about our shared cultural past and the development of human civilisation, but our understanding of early games is incomplete and often based on unreliable reconstructions. 
This paper describes the Digital Ludeme Project, a five-year research project currently underway that aims to address such issues using modern computational techniques. 
\end{abstract}

\begin{IEEEkeywords}
ancient games, ludemes, general game playing, phylogenetics, history of mathematics, digital archaeoludology
\end{IEEEkeywords}


\section{Introduction}

The development of games has gone hand-in-hand with the development of human civilisation~\cite{Huizinga1938}. 
However, our knowledge of early games is incomplete and based on often unreliable interpretations of available evidence. 

There has been a wealth of traditional game studies over recent centuries, from historical, anthropological, arch{\ae}ological -- and more recently ethnological and cultural -- perspectives. 
There is now also a wealth of computational game studies; games have always been a driving factor behind {\it artificial intelligence} (AI) and {\it machine learning} (ML) research since the inception of these fields in the 1950s~\cite{Brooks2017}, and now Game AI is maturing as a research field in its own right~\cite{Yannakakis2018}. 

However, there has been little overlap between computational and historical studies of traditional games. 
This paper outlines a newly launched research project aimed to address this gap, so that our historical understanding of games might benefit from current advances in technology. 


\subsection{The Digital Ludeme Project}

The Digital Ludeme Project\footnote{http://www.ludeme.eu} is a five-year research project being conducted at Maastricht University over 2018--23, funded by a European Research Council (ERC) Consolidator Grant.
The objectives of the project are to:

\begin{enumerate}

\item {\it Model} the full range of traditional strategy games in a single, playable digital database.

\item {\it Reconstruct} missing knowledge about traditional strategy games with an unprecedented degree of accuracy. 

\item {\it Map} the development of traditional strategy games and explore their role in the development of human culture and the spread of mathematical ideas. 

\end{enumerate}


An ultimate goal of the project is to produce a ``family tree'' of the world's traditional strategy games, with which the dispersal of games and related mathematical ideas might be traced throughout recorded history. 
This will pioneer a new field of study called {\it digital arch{\ae}oludology} (DA) which will involve the use of modern computational techniques for the analysis and reconstruction of traditional games from incomplete descriptions, to provide new tools and techniques for improving our understanding of their development. %

This paper describes the project's research context, scope of games to be covered, the methodology used, and some potential benefits. 


\section{Research Context}

While there is much arch{\ae}ological evidence of ancient games, the rules for playing them are usually lost~\cite{Murray1952} and must be reconstructed by modern historians according to their knowledge of the cultures in which they were played~\cite{Schadler1998,Schadler2013}. 
The rules for ancient and early games were typically passed on through oral tradition rather than being transcribed, which may have contributed to their variation and embellishment into the range of games that we see today~\cite{Murray1952}, but means that our understanding of early games is largely based on modern reconstructions. 
The following examples demonstrate some of the issues involved.

\begin{figure}[htbp]
\centerline{\includegraphics[width=1\columnwidth]{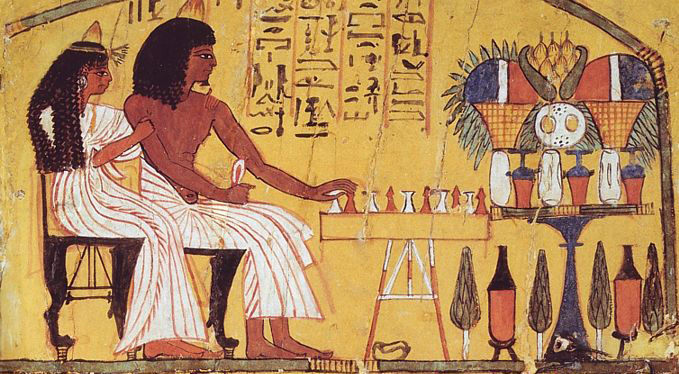}}
\caption{Ancient Egyptian hieroglyphic art showing Senet being played.}
\label{fig:Senet}
\end{figure}


\subsection{Missing Knowledge}

Many boards and sets of playing pieces have been found for the ancient Egyptian game of Senet dating back to c.3500{\sc bc}, some in pristine condition, allowing historians to deduce with reasonable certainty what type of game it was~\cite{Crist2016}. 
However, the only known clues as to {\it how} Senet was played are found in hieroglyphic art dating back to c.3100{\sc bc}, such as Fig.~\ref{fig:Senet}), which shows stylised characters playing the game. 

Game historian H.J.R. Murray declined to propose a complete set of rules for Senet in his classic 1952 book~\cite{Murray1952}. 
Kendall did so in 1978~\cite{Kendall1978}, but his version is based on snippets of information from sources spanning thousands of years and remains questionable. 
For example, Murray interpreted board squares marked with certain sacred Egyptian symbols as entry points for pieces, while Kendall interpreted these as points of departure from the game~\cite{Duggan2015}.


\subsection{Loss of Knowledge}

Even when records of the rules for ancient games {\it are} found, interpretation can be problematic. 
For example, the earliest known record of a game's rule set is for the Royal Game of Ur. 
Game sets were uncovered in Iraq dating back to 2600--2400{\sc bc}, but it was not until Irving Finkel's 1990 study of two Sumerian stone tablets dating from 177--176{\sc bc}~\cite{Finkel2007} that the game's (probable) rules were found. 

\begin{figure}[htbp]
\centerline{\includegraphics[width=1\columnwidth]{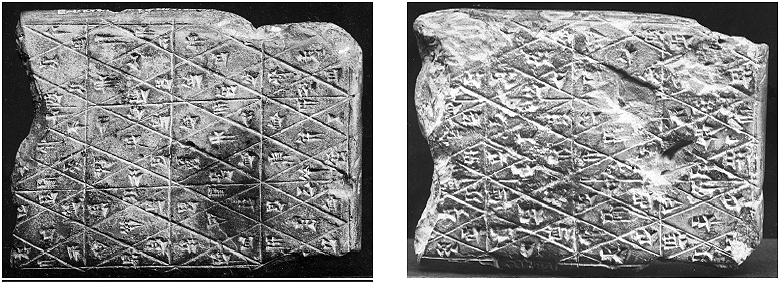}}
\caption{Sumerian tablet showing game rules destroyed in World War II.}
\label{fig:Ur}
\end{figure}

The first tablet was found by Finkel among 130,000 such tablets held in the British Museum. 
The second tablet (Fig.~\ref{fig:Ur}) was luckily photographed shortly before its destruction in a Parisian studio during World War II and recognised by Finkel half a century later~\cite{Finkel2007}. 
Note that Finkel's interpretation of the game was made thousands of years after the tablets were made, which was itself thousands of years after the game was originally played. 

In addition to wanton destruction caused by war, vandalism, looting, desecration, etc., more benign forces such as erosion and urban development can also take their toll on arch{\ae}ological evidence. 
For example, Crist describes the case of ancient game boards etched into rock surfaces in Azerbaijan~\cite{Crist2018}, which were destroyed to make way for a housing development between one research trip and the next. 


\subsection{Translation Errors}

\begin{figure}[htbp]
\centerline{\includegraphics[width=1\columnwidth]{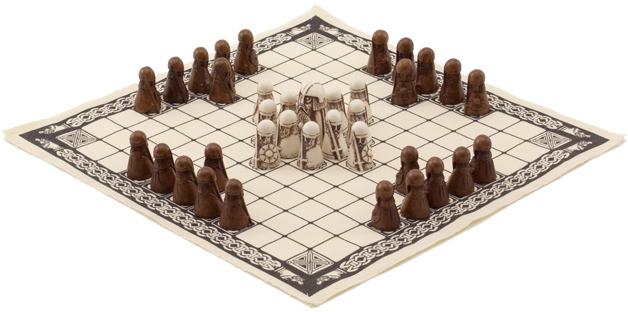}}
\caption{A modern version of the Viking game Hnefatafl.}
\label{fig:Hnefatafl}
\end{figure}

Translation errors are another issue. 
Consider the game of Hnefatafl, played by Vikings from c.400{\sc bc} and spread wherever they travelled, for which no known documentation of the original rules exists. 
A modern set is shown in Fig.~\ref{fig:Hnefatafl} 

In 1732, Swedish naturalist Carolus Linn{\ae}us observed the related game Tablut being played in Lappland, and recorded its rules in Latin in his travel diary~\cite{Linnaeus1811}. 
This account was translated by E. L. Smith in 1811, who mistook the phrase ``likewise the king'' to mean ``except the king'', to produce a biased rule set that greatly favoured the king's side~\cite{Ashton2010}. 
Murray used this translation as the basis for his 1913 reconstruction of Hnefatafl~\cite{Murray1913}, making it the definitive rule set for many years~\cite{Murray1952}, until players and researchers subsequently corrected this flaw to give the many versions of the game played today.


\subsection{Reconstruction Errors}

\begin{figure}[htbp]
\centerline{\includegraphics[width=1\columnwidth]{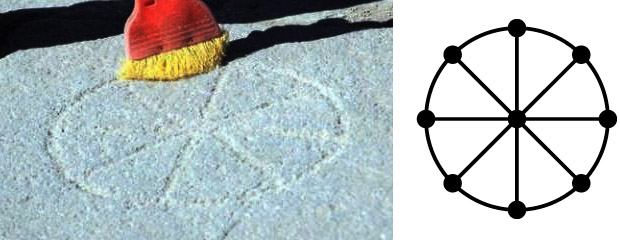}}
\caption{Engraving found at Assos (Turkey) and a Small Merels board.}
\label{fig:Assos}
\end{figure}

Often the arch{\ae}ological evidence of games provides little clue as to their rules, which must be deduced almost entirely from context. 
For example, the rock etching shown in Fig.~\ref{fig:Assos} (left) was unearthed in Assos, Turkey, and estimated to be around 2,300 years of age~\cite{Ertugrul2015}. 
This design (right) is listed as design \#88 in Uberti's census of Merels 
boards~\cite{Uberti2015}, which revealed around 100 designs from more than 2,500 examples across 43 countries. 

It is assumed that this board was used to play Round Merels,\footnote{A miniature version of the traditional Merels or Nine Men's Morris.} which seems to be the default assumption for boards of this design, location and epoch. 
German historian Carl Bl{\"u}mlein proposed a plausible reconstruction of the rules in 1918, which became the accepted standard and was not questioned for almost 100 years, when a 2014 analysis revealed a critical flaw that allowed players to exploit infinite cycles~\cite{Heimann2014}. 
It is now questioned whether this design was used for a different type of game or was not even a game at all.  


\subsection{Transcription Errors}

Transcription errors can be an issue even with the records of more recent games. 
For example, the design shown in Fig.~\ref{fig:Assos} (right) -- a wheel with eight spokes -- is also used for the 18\textsuperscript{th} century Maori game Mu Torere from New Zealand, even though its rules are quite different to Round Merels. 

\begin{figure}[htbp]
\centerline{\includegraphics[width=1\columnwidth]{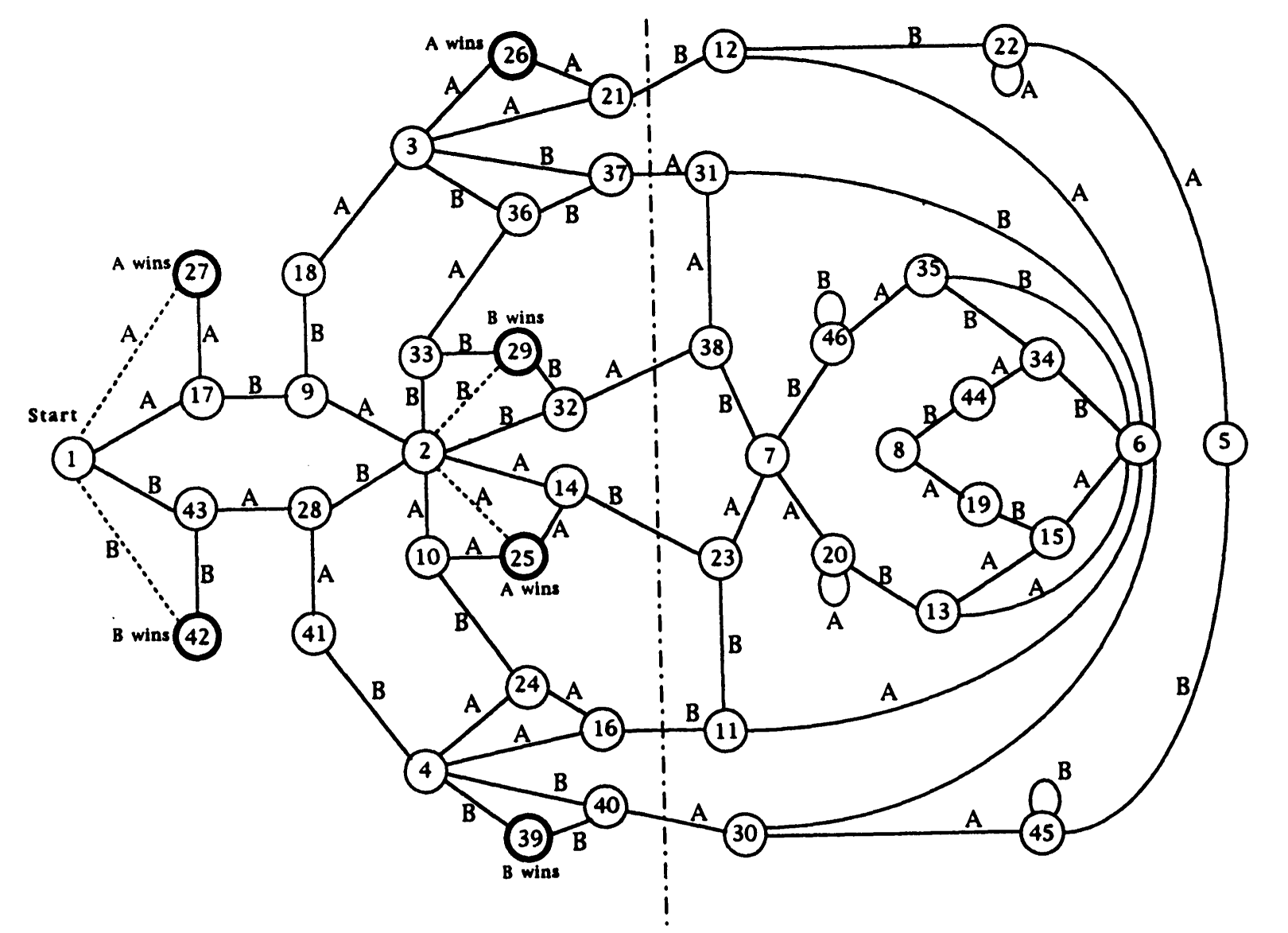}}
\caption{Full game tree expansion of Mu Torere showing trivial wins (dotted).}
\label{fig:GameTree}
\end{figure}

Ethnomathematician Marcia Ascher noted in her 1987 study of Mu Torere~\cite{Ascher1987} that at least two historians' transcriptions of its rules simplified out an apparently trivial starting rule, without which the game became unplayable. 
A full game tree expansion by mathematician Philip Straffin~\cite{Straffin1995} (Fig.~\ref{fig:GameTree}) shows that either player can win on the first move if this rule is not used, 
which is obvious from even the simplest analysis. 

\subsection{Reinvention Estimates}

\begin{figure}[htbp]
\centerline{\includegraphics[width=1\columnwidth]{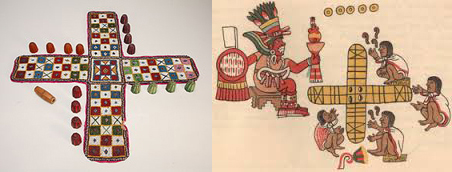}}
\caption{Similar boards for Pachisi (left) and Patolli (right).}
\label{fig:PachisiPatolli}
\end{figure}

The discovery of similar game boards in India and ancient Mexico -- Pachisi and Patolli (Fig.~\ref{fig:PachisiPatolli}) -- was used in 1879 as evidence of early pre-Columbian contact between Asia and South America~\cite{Tylor1879}, even though the rules for each game are quite different~\cite{Caso1925}. 
This claim was disputed half a century later due to the notion of ``limited possibilities'' in design making coincidental reinvention more plausible~\cite{Erasmus1950}, even though Murray~\cite{Murray1952} points out that such independent reinvention is generally unlikely. 

Which view is more likely to be correct? 
Such analyses will remain speculation until methods are developed to provide quantitative evidence for such cases.
A more accurate and complete knowledge of the development and spread of traditional games could help clarify such cases, and shed new light on trade routes and points of contact between cultures. 


\subsection{Partial Evidence}

\begin{figure}[htbp]
\centerline{\includegraphics[width=1\columnwidth]{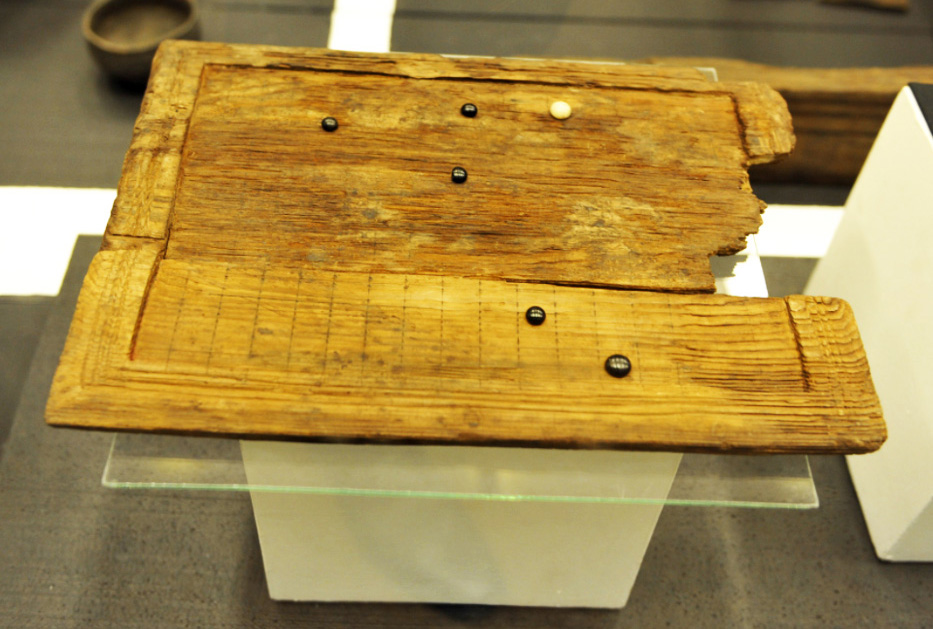}}
\caption{Partial game set found in a Slovakian tomb.}
\label{fig:Poprad}
\end{figure}

A challenging task facing historians is to reconstruct the rules of games when only some of the equipment is known. 
For example, Fig.~\ref{fig:Poprad} shows a game board and pieces dated to 375{\sc ad} and found in 2006 in Poprad, Slovakia, in the tomb of a Germanic chieftain who served in the Roman army~\cite{Spectator2018}. 
This equipment has no clear precedent in Europe, and historian Ulrich Sch\"{a}dler describes the reconstruction of the game's original rules as ``impossible'', as the board is incomplete and only a few playing pieces survived.\footnote{Conversation at the Board Game Studies Colloquium, Athens, April 2018.}

%
%


These examples -- and many others from around the world -- highlight the difficulty of compiling reliable knowledge of traditional games. 
What little evidence does exist is fragile and easily lost, 
and attempts to reconstruct missing rule sets have so far relied heavily on historical context rather than mathematical evaluation, but once accepted into the canon become the {\it de facto} standards.  
Thus, our knowledge of this important part of our cultural heritage is at best partial, or skewed by unreliable reconstructions.  

While much attention has been paid to ensuring the historical {\it authenticity} of reconstructions, there has been to date no systematic approach to evaluating the {\it quality} of proposed reconstructions as games. 
This project aims to develop tools and methods for improving our understanding of traditional games with unprecedented mathematical rigour.


\section{Scope}    \label{sec:Scope}

The Digital Ludeme Project deals with {\it traditional games of strategy}, i.e. games with no proprietary owner~\cite[p.5]{Parlett1999} that exist in the pubic domain,\footnote{The more precise distinction between traditional games and those invented by known individuals and distributed by games companies~\cite{Horn2008} can lead to ambiguous cases~\cite{deVoogt2018}.} and in which players succeed through mental rather than physical acumen. 
This category includes most board games, some card games, some dice games, some tile games, etc., and may involve non-deterministic elements of chance or hidden information as long as strategic play is rewarded over random play. 
It excludes dexterity games, social games, sports, video games, etc.

\begin{figure}[htbp]
\centerline{\includegraphics[width=1\columnwidth]{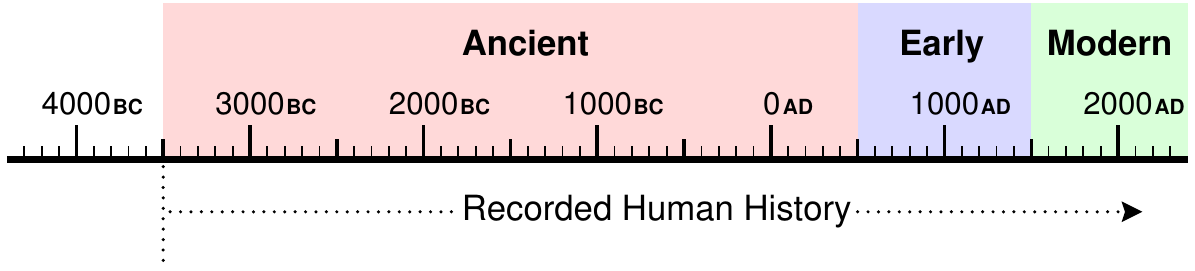}}
\caption{Timeline of key periods in recorded human history.}
\label{fig:Timeline}
\end{figure}

This study will cover the full range of traditional strategy games throughout recorded human history, i.e. from around 3500{\sc bc}, from all countries and cultures worldwide. 
Within this context, it is useful to distinguish the approximate time periods shown in Fig.~\ref{fig:Timeline}:

\begin{itemize}

\item {\it Ancient}: before 500{\sc ad}.

\item {\it Early}: 500{\sc ad} -- 1500{\sc ad}.

\item {\it Modern}: after 1500{\sc ad}.

\end{itemize}

In general, the older a game is, the less is known about it. 
The original rules are known for most modern games, some early  games, but few ancient games. 


\subsection{Influencers}

It would be unrealistic to attempt to model {\it every} traditional strategy game. 
For example, of the thousands of known Chess variants,\footnote{http://
www.chessvariants.com} hundreds could fall under the umbrella of ``traditional''. 
There also exist over 800 known variants of Mancala, let alone the undocumented ones~\cite{deVoogt1999}. 

It is difficult to even estimate the number of known traditional strategy games. 
The BoardGameGeek (BGG) online database\footnote{http://www.boardgamegeek.com} lists around 100,000 known board games with (probable) invention dates and details regarding designer and publisher. 
However, entire families of games, such as the hundreds of Mancala variants, are typically collapsed into a few representative entries, making the total number of BGG entries a gross underestimate of the number of actual games, possibly by orders of magnitude. 

Instead, the Digital Ludeme Project will investigate a representative sample of 1,000 of the world's traditional strategy games, which include the most influential examples throughout history. 
Such {\it influencers} might be identified by:

\begin{itemize}

\item {\it Appeal:} Estimated total number of players.

\item {\it Impact:} Number of similar games that follow.

\item {\it Importance:} Footprint in the literature.

\end{itemize}

The idea is to focus on those games that are most important to the evolution of traditional strategy games. 
Games that are known to have existed, but which might have only been played within one community or even one family, and for which there is no evidence of any influence over later games, constitute evolutionary dead ends that are of less interest for this task. 


\section{Ludemes}

Games will be modelled as structures of {\it ludemes}, i.e. game memes or conceptual units of game-related information~\cite{Parlett2016}. 
These constitute a game's underlying building blocks, and are the high-level conceptual terms that human designers use to understand and describe games. 
Previous work on evolving board games~\cite{Browne2011a} demonstrated the effectiveness of the ludemic model for the automated generation of games.



Table~\ref{tab:Ludemic} shows how the game of Tic-Tac-Toe might be described in ludemic form. 
This description is simple, clear, and encapsulates key concepts and labels them with meaningful names. 
Breaking games down into ludemes makes them easier to model, compare and manipulate digitally, and makes it possible the model the full range of traditional games in a single playable database. 



\begin{table}[htbp]
\caption{Ludemic Description of Tic-Tac-Toe}
\begin{center}
\begin{tabular} {| l | }
\hline
\\
{\tt \footnotesize (game Tic-Tac-Toe} \\
{\tt \footnotesize \ \ \ \ (players White Black)} \\
{\tt \footnotesize \ \ \ \ (equipment} \\
{\tt \footnotesize \ \ \ \ \ \ \ \ (board (square 3) diagonals)} \\
{\tt \footnotesize \ \ \ \ )} \\
{\tt \footnotesize \ \ \ \ (rules} \\
{\tt \footnotesize \ \ \ \ \ \ \ \ (play (add (piece Own) (board Empty)))} \\
{\tt \footnotesize \ \ \ \ \ \ \ \ (end (win All (line 3 Own Any)))} \\
{\tt \footnotesize \ \ \ \ )} \\
{\tt \footnotesize )} \\
\\
\hline
\end{tabular}
\label{tab:Ludemic}
\end{center}
\end{table}


\subsection{Stanford GDL}

Table~\ref{tab:GDL} shows the same game described in the Stanford Logic Group's {\it game description language} (GDL), which has become the standard method for describing games in {\it general game playing} (GGP) research~\cite{Genesereth2005}.
GDL offers benefits of transparency (the game description itself contains the instructions for updating the game state) and  correctness checking (that the rules are well-formed). 

\begin{table}[htbp]
\caption{GDL Description of Tic-Tac-Toe}
\begin{center}
\begin{tabular} {| l | }
\hline 
\\
{\tt \tiny (role white)} \\
{\tt \tiny (role black)} \\
{\tt \tiny (init (cell 1 1 b))} \\
{\tt \tiny (init (cell 1 2 b))} \\
{\tt \tiny (init (cell 1 3 b))} \\
{\tt \tiny (init (cell 2 1 b))} \\
{\tt \tiny (init (cell 2 2 b))} \\
{\tt \tiny (init (cell 2 3 b))} \\
{\tt \tiny (init (cell 3 1 b))} \\
{\tt \tiny (init (cell 3 2 b))} \\
{\tt \tiny (init (cell 3 3 b))} \\
{\tt \tiny (init (control white))} \\
{\tt \tiny (<= (legal ?w (mark ?x ?y))(true (cell ?x ?y b)) (true (control ?w)))} \\
{\tt \tiny (<= (legal white noop) (true (control black)))} \\
{\tt \tiny (<= (legal black noop) (true (control white)))} \\
{\tt \tiny (<= (next (cell ?m ?n x)) (does white (mark ?m ?n)) (true (cell ?m ?n b)))} \\
{\tt \tiny (<= (next (cell ?m ?n o)) (does black (mark ?m ?n)) (true (cell ?m ?n b)))} \\
{\tt \tiny (<= (next (cell ?m ?n ?w))(true (cell ?m ?n ?w)) (distinct ?w b))} \\
{\tt \tiny (<= (next (cell ?m ?n b)) (does ?w (mark ?j ?k)) (true (cell ?m ?n b)) } \\ 
{\tt \tiny \ \ \ \  (or (distinct ?m ?j) (distinct ?n ?k)))} \\
{\tt \tiny (<= (next (control white))(true (control black)))} \\
{\tt \tiny (<= (next (control black))(true (control white)))} \\
{\tt \tiny (<= (row ?m ?x)(true (cell ?m 1 ?x))(true (cell ?m 2 ?x))(true (cell ?m 3 ?x)))} \\
{\tt \tiny (<= (column ?n ?x)(true(cell 1 ?n ?x))(true(cell 2 ?n ?x))(true(cell 3 ?n ?x)))} \\
{\tt \tiny (<= (diagonal ?x)(true (cell 1 1 ?x))(true (cell 2 2 ?x))(true (cell 3 3 ?x)))} \\
{\tt \tiny (<= (diagonal ?x)(true (cell 1 3 ?x))(true (cell 2 2 ?x))(true (cell 3 1 ?x)))} \\
{\tt \tiny (<= (line ?x) (row ?m ?x))} \\
{\tt \tiny (<= (line ?x) (column ?m ?x))} \\
{\tt \tiny (<= (line ?x) (diagonal ?x))} \\
{\tt \tiny (<= open (true (cell ?m ?n b))) (<= (goal white 100) (line x))} \\
{\tt \tiny (<= (goal white 50) (not open) (not (line x)) (not (line o)))} \\
{\tt \tiny (<= (goal white 0) open (not (line x)))} \\
{\tt \tiny (<= (goal black 100) (line o))} \\
{\tt \tiny (<= (goal black 50) (not open) (not (line x)) (not (line o)))} \\
{\tt \tiny (<= (goal black 0) open (not (line o)))} \\
{\tt \tiny (<= terminal (line x))} \\
{\tt \tiny (<= terminal (line o))} \\
{\tt \tiny (<= terminal (not open)) } \\
\\
\hline
\end{tabular}
\label{tab:GDL}
\end{center}
\end{table}

By contrast, the ludemic approach hides the implementation details to provide simplicity, encapsulation, efficiency and ease of use. 
For example, if we wish to modify Tic-Tac-Toe so that players aim to make a line-of-4 on a 5$\times$5 board, or play on a hexagonal grid, or aim for some other winning condition altogether, then each of these changes would involve a trivial parameter adjustment or swapping of keywords in ludemic form. 
Implementing these changes in GDL format, however, would require a significant amount of the code to be rewritten and retested.
The encapsulation of concepts makes ludemic descriptions easier to modify and more evolvable than GDL descriptions.


\section{L{\sc udii} System}

A complete {\it general game system} (GGS) for modelling, playing, analysing, optimising and generating the full range of traditional strategy games is being developed for this project. 
This system, called {\sc Ludii}, is based on similar principles to the previous {\sc Ludi} system~\cite{Browne2009}, but improved in almost every way to be more general, extensible and efficient. 

The core of {\sc Ludii} is a {\it ludeme library} consisting of a number of Java classes each implementing a particular ludeme. 
Games are described as structured sets of ludemes, as per Table~\ref{tab:Ludemic}, according to an EBNF-style grammar automatically generated from the ludeme library's class hierarchy using a {\it class grammar} approach~\cite{Browne2016}. 
Game descriptions can then be compiled directly to Java byte code according to their underlying ludeme classes. 

\subsection{Plausible AI}

AI move planning will be performed using {\it Monte Carlo tree search} (MCTS)~\cite{Browne2012} with playouts biased by strategies learnt through self-play. 
MCTS has become the preferred approach for general game playing over recent years, due to its ability to devise plausible actions in the absence of any tactical or strategic knowledge about the given task. 
Although it can prove weaker for some games than others, it provides a good baseline level of AI play for most games. 

The combination of deep learning with MCTS has recently had spectacular success with Go~\cite{Silver2016}. 
However, this level of superhuman performance is not required for this project, 
where a more modest level of play pitched just beyond average human level is preferable, in order to estimate the potential of games to interest human players. 
Superhuman AI that plays differently to humans could actually bias evaluations;  
instead, we want an AI that makes moves that human players would plausibly make. 


\subsubsection{Lightweight Local Features}

\begin{figure}[htbp]
\centerline{\includegraphics[width=.75\columnwidth]{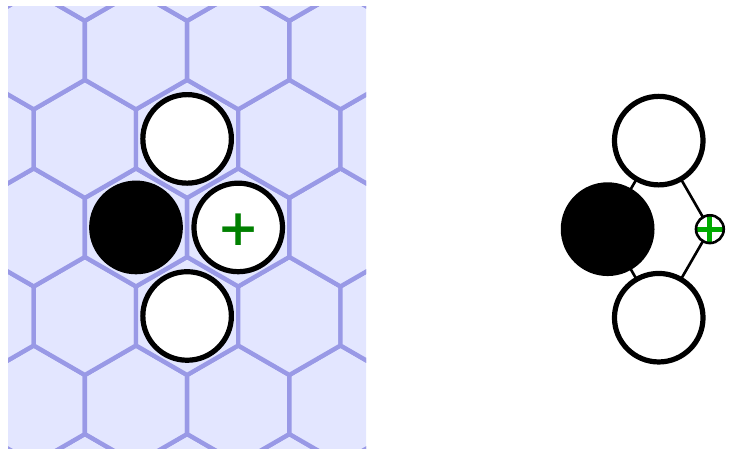}}
\caption{A strong pattern for connection games on the hexagonal grid.}
\label{fig:HexPatterns}
\end{figure}

To elevate MCTS to a sufficient level of play for all games, playouts will be biased with domain-dependent information in the form of lightweight features that capture geometric piece patterns, learnt through self-play. 
For example, the pattern shown in Fig.~\ref{fig:HexPatterns}, which completes a threatened connection in connection games played on the hexagonal grid, improves MCTS playing strength when incorporated into the playouts of such games~\cite{Raiko2008}.  

Such patterns represent local strategies that human players typically learn to apply. 
They will not capture more complex global strategies, but should serve to improve MCTS to plausible levels of play, and -- importantly -- could give an indication of a game's strategic potential. 


\subsection{Game Evaluation}

When evaluating rule sets, it is important to consider the {\it quality} of the resulting games, which is the vital element missing from many historical studies of games. 
If a rule set is flawed, or does not have potential to interest human players, then it is unlikely that is how the game was actually played.  

Previous work~\cite{Althofer2003, Browne2009} has outlined robust indicators of flaws in games that can be easily measured through self-play:

\begin{enumerate}

\item {\it Length:} Games should not be too short or long.

\item {\it Fairness:} Games should not unduly favour either player.

\item {\it Drawishness:} Games should not end in draws too often.

\end{enumerate}

The question of what makes a game ``good'' in players' eyes is much more difficult; there are no universal indicators of game quality, and preferences can differ between individuals, cultures, and across time. 
However, it would make sense that a key quality for strategy games should be their {\it strategic depth}, indicated by the number and complexity of potential strategies that players can learn. 

\begin{figure}[htbp]
\centerline{\includegraphics[width=.85\columnwidth]{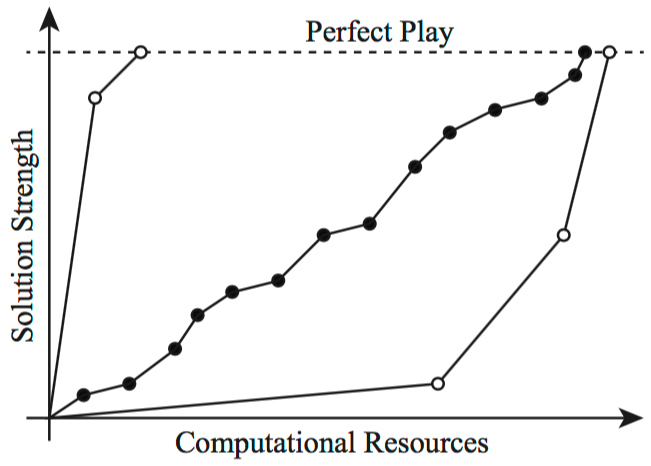}}
\caption{Linear acquisition of strategies in an interesting game (black)~\cite{Lantz2017}.}
\label{fig:StrategyLadder}
\end{figure}

Lantz {\it et al.} propose the notion of the {\it strategy ladder}~\cite{Lantz2017}. 
Fig.~\ref{fig:StrategyLadder} shows three plots that represent three different games, with dots indicating relevant strategies that players can learn.
The leftmost game (white dots) is uninteresting as it has a simple winning strategy that is easy to learn. 
The rightmost game (white dots) is uninteresting as it has difficult strategies that are too hard to learn. 
The middle game (black dots) has a variety of strategies of linearly increasing complexity; the player can immediately see some simple strategies, but learn more complex strategies as the game is played more.  


Strategic depth should be considered relative to a game's complexity. 
For example, Mu Torere (46 legal positions~\cite{Straffin1995}) could be expected to involve fewer strategies than Go ($\approx$$2.08$$\times$$10^{170}$ legal positions~\cite{Tromp2016}). 
Other ways to estimate the strategic depth of a game might include comparing relative win rates over a  range of AI agents of varying strength. 


\subsection{Strategy Learning, Transfer and Explanation}

If these lightweight features based on piece patterns represent local strategies, then the number and complexity of learnt features could give an indication of a game's strategic depth. 
Basing local piece patterns on the adjacency of a game's underlying graph (Fig.~\ref{fig:HexPatterns}, right) rather than the board itself (Fig.~\ref{fig:HexPatterns}, left) provides geometric independence that allows learnt features to be transferred between different board types. 
Keeping the feature attributes as simple and abstract as possible makes it more likely that features might also be transferrable to other game types. 

\begin{figure}[htbp]
\centerline{\includegraphics[width=.75\columnwidth]{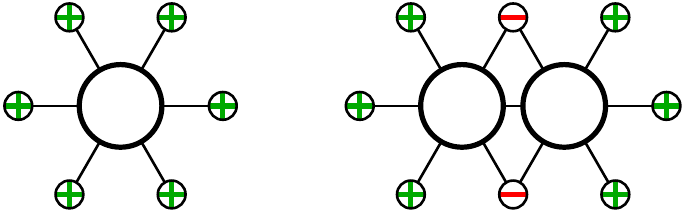}}
\caption{Patterns that constitute a ``make thin groups'' strategy.}
\label{fig:QL}
\end{figure}

The fact that ludemes are labelled with meaningful names raises the possibility of automatically explaining learnt strategies in human-comprehensible terms. 
For example, the strategy encoded in Fig.~\ref{fig:HexPatterns} might be explained as ``complete threatened connections between your pieces''. 
The strategy encoded in Fig.~\ref{fig:QL}, effective for the recent game Quantum Leap~\cite{Browne2014b}, might be explained as ``make thin groups of your pieces'', by encouraging the growth of singletons and the extension of adjacent pieces except at mutually adjacent points. 

\section{Genetics of Games}

In order to map the dispersal 
of traditional strategy games, it is useful to cast the mechanism for their evolution into a biological genetic  framework. 
Anthropologist Alex de Voogt has stated:
{\it There is nothing genetic about board games. 
There are no genes or mental parameters that only change with a new generation of people as in linguistics or in biology}~\cite[p.105]{deVoogt1999}.
However, I would argue that the ludemic model allows us to distinguish between the {\it form} of a game defined by its ludemic makeup of rules and equipment (i.e. genotype) and the {\it function} of a game defined by the behaviour it exhibits when played (i.e. phenotype). 
Ludemes are the ``DNA'' that define each game, and the ludemic approach has already proven to be a valid and powerful model for evolving games~\cite{Browne2009}.


\subsection{Computational Phylogenetics}

Once a genetic framework has been established, {\it computational phylogenetics} techniques such as those used to create phylogenetic trees mapping the dispersal of human language~\cite{Greenhill2015} can be applied. 
Such techniques allow {\it ancestral state reconstruction} for estimating the likelihood of given traits occurring in ``ancestor'' games, and the inference of possible {\it missing links} in the form of unknown games suggested by the phylogenetic record for which no evidence exists. 

Phylogenetic techniques have previously been applied to subsets of Mancala games~\cite{Eagle1999} and Chess-like games~\cite{Kraaijeveld2001}. 
However, phylogenetic analyses of such cultural domains 
tend to confuse the genotype and phenotype of artefacts, yielding classifications of questionable value 
based on superficial traits rather than meaningful underlying structures~\cite{Morrison2013}. 
List {\it et al.} provide guidelines for correctly casting cultural domains in a biological framework~\cite{List2016}.

\subsection{Game Distance}

Games do not contain the traces of genetic heritage that biological organisms do; rule sets are typically optimised and superfluous rules stripped out, making their heritage hard to trace.
In lieu of a metric for genetic distance, the {\it ludemic distance} between games will be used, given by the {\it weighted edit distance} (WED) between ludemic descriptions, i.e. the number of removals, insertions and edits required to convert one into the other, weighted according to the relative importance of each attribute. 
This is similar in principle to the {\it Hamming distance} used to quantify the similarity between DNA sequences in bioinformatics~\cite{Anselmo2012}. 
Care must be taken to detect and handle {\it homologies}~\cite{Temken2007} that occur when different ludeme structures produce the same behaviour in play. 


\subsection{Horizontal Influence Maps}

\begin{figure}[htbp]
\centerline{\includegraphics[width=.85\columnwidth]{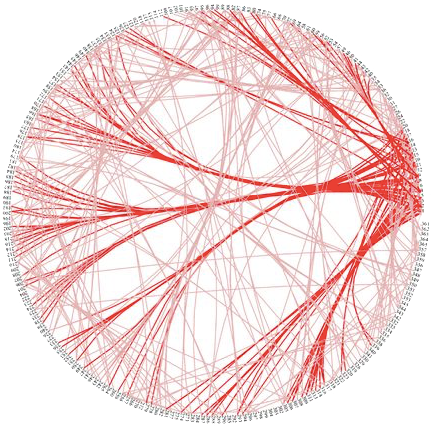}}
\caption{Horizontal influence map (from~\cite{Valverde2016}).}
\label{fig:HIM}
\end{figure}

Morrison points out that phylogenetic {\it networks} may be more suitable than {\it trees} for modelling the evolution of cultural artefacts~\cite{Morrison2014}. 
This seems especially relevant for games, which are more likely to have evolved through distributed {\it polygenesis} from multiple sources  than {\it monogenesis} from a single common ancestor~\cite{Parlett2011}, and in which rules can pass from one to another through {\it ethnogenesis} (i.e. horizontal gene transfer) rather than classic inheritance. 
The prevalence of ethnogenesis in the spread of games could warrant the use of {\it horizontal influence maps} (HIM)~\cite{Valverde2016} 
(Fig.~\ref{fig:HIM}) rather than standard phylogenetic approaches based on vertical gene transfer.  

\section{Cultural Mapping of Games}

To facilitate the cultural mapping of games, ludemes and game descriptions will be tagged with relevant metadata:

\begin{itemize}

\item {\it Mathematical:} Ludeme classes will be tagged with the underlying mathematical concepts that they embody.

\item {\it Historical:} Game descriptions will be tagged with details regarding when and where they were played (among other cultural details). 

\end{itemize}

Each game will therefore have a {\it mathematical profile} based upon its component ludemes and a {\it historical profile}. 
The game database will be data-mined for common ludemeplexes that represent important game mechanisms.  
The associated metadata will be cross-referenced to create  
{\it knowledge graphs} that give probabilistic models~\cite{Dong2014} of the relationships between their geographical, historical and mathematical dimensions. 

The cultural location of games will be achieved using a geo-location service such as GeaCron.\footnote{http://geacron.com}
GeaCron maintains a database of geo-political world maps for every year from 3000{\sc bc} to the present day, which can be queried to specify which empire, nation, civilisation or culture was dominant at any given geographical location in recorded history. 

\begin{figure}[htbp]
\centerline{\includegraphics[width=1\columnwidth]{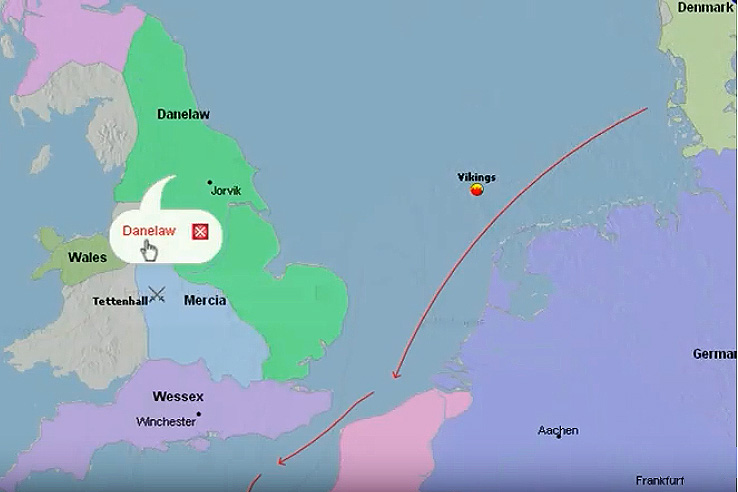}}
\caption{Viking route from Denmark to Paris in 845{\sc ad} (image by GeaCron).}
\label{fig:GeaCron}
\end{figure}

GeaCron also provides details of known trade routes, expeditions, and other key historical events, for example Fig.~\ref{fig:GeaCron} shows the Viking route from Denmark to Paris in 845{\sc ad}. 
This provides a mechanism for correlating the spread of games, ludemes and associated mathematical ideas with the spread of human civilisation. 



\section{Digital Arch{\ae}oludology}

With these ideas in mind, I propose a new field of study called {\it digital arch{\ae}oludology} (DA), for the analysis and reconstruction of ancient games from incomplete descriptions using modern computational techniques. 
The aim is to provide tools and methods that might help game historians and researchers better understand traditional games.



Traditional game studies have tended to focus on the authenticity of reconstructions (as cultural artefacts) rather than their actual quality as games. 
DA seeks to redress this imbalance by searching for plausible reconstructions that maximise both quality and historical authenticity, hopefully leading to better reconstructions, a better understanding of ancient and early games, and a more accurate and complete model of the development of traditional strategy games throughout history. 

%
%
%
%


\subsection{Forensic Game Reconstruction}

A key application of DA is the forensic reconstruction of games from partial descriptions, such as the Poprad game shown in Fig.~\ref{fig:Poprad}. 
The following equipment is known:

\begin{enumerate}

\item Rectangular board with 17$\times$15 or 17$\times$16 square grid.

\item Pieces of two colours.

\item Pieces of two sizes (possibly). 

\end{enumerate}

The {\sc Ludii} system could perform a search of the ludeme space constrained to these requirements, to find plausible rule sets that maximise both game quality and historical authenticity based on what is known about the game, in this case the historical and cultural context of the tomb in which the game was found and its inhabitant. 
{\sc Ludii} could fill in the ``missing bits'' such as finding historically accurate combinations of rules that provide interesting games, what number of pieces provide better results, how they are best arranged to start the game, and so on. 
The aim is to provide tools for the plausible reconstruction of such missing knowledge, so such tasks no longer seem impossible. 

Reviewing the complete set of traditional game reconstructions modelled in the database -- to identify implausible cases and optimise them where possible -- has the potential to improve our understanding of traditional games. 
The intention is to create a positive feedback loop in which better reconstructions lead to better historical and cultural mappings, which lead to even better reconstructions, and so on.

\section{Conclusion}

Games offer a rich window of insight into our cultural past, but early examples were rarely documented and our understanding of them is incomplete. While there has been considerable historical research into games and their use as tools of cultural analysis, much is based on the interpretation of partial evidence with little mathematical analysis. 
This project will use modern computational techniques to help fill these gaps in our knowledge empirically, establishing a new field of research called digital arch{\ae}oludology.


\section*{Acknowledgment}

This work is part of the Digital Ludeme Project, funded by \euro2m European Research Council (ERC) Consolidator Grant \#771292, conducted at Maastricht University over 2018--23.




\end{document}